# Absorption and emission modulation in MoS$_2$-GaN (0001) heterostructure by interface phonon-exciton coupling

**YUBA POUDEL[1], JAGODA SLAWINSKA[1], PRIYA GOPAL[1], SAIRAMAN SEETHARAMAN[2], ZACHARIAH HENNIGHAUSEN[3], SWASTIK KAR[3], FRANCIS D'SOUZA[2], MARCO BUONGIORNO NARDELLI[1], ARUP NEOGI[1,*]**

[1]*Department of Physics, University of North Texas, Denton Texas, 76203, USA*
[2]*Department of Chemistry, University of North Texas, Denton Texas, 76203, USA*
[3]*Department of Physics, Northeastern University, Boston, Massachusetts 02115, USA*
*\* Corresponding author Email: arup@unt.edu*

**Semiconductor heterostructures based on layered two-dimensional transition metal dichalcogenides (TMD) interfaced to gallium nitride (GaN) are excellent material systems to realize broadband light emitters and absorbers. The surface properties of the polar semiconductor, such as GaN are dominated by interface phonons, thus the optical properties of the vertical heterostructure depend strongly on the interface exciton-phonon coupling. The origin and activation of different Raman modes in the heterostructure due to coupling between interfacial phonons and optically generated carriers in a monolayer MoS$_2$-GaN (0001) heterostructure was observed. This coupling strongly influences the non-equilibrium absorption properties of MoS$_2$ and the emission properties of both semiconductors. Density functional theory (DFT) calculations were performed to study the band alignment of the interface, which revealed a type-I heterostructure. The optical excitation with interband transition in MoS$_2$ at K-point strongly modulates the C excitonic band in MoS$_2$. The overlap of absorption and emission bands of GaN with the absorption bands of MoS$_2$ induces the energy and charge transfer across the interface with an optical excitation at Γ-point. A strong modulation of the excitonic absorption states is observed in MoS$_2$ on GaN substrate with transient optical pump-probe spectroscopy. The interaction of carriers with phonons and defect states leads to the enhanced and blue shifted emission in MoS$_2$ on GaN substrate. Our results demonstrate the relevance of interface coupling between phonons and carriers for the development of optical and electronic applications.**

## 1. INTRODUCTION

Gallium nitride (GaN) is extensively studied for the optoelectronic applications such as light emitters, high electron mobility transistors and photodetectors [1]–[3]. The broadband light emitters can be constructed based on ternary compounds such as indium gallium nitride (InGaN) formed from III-V bulk semiconductors, which offer tunable band gaps from 3.4 eV (GaN) to 0.64 eV (InN) [4], [5]. However, a high In content required to generate red or near-infrared light decreases the efficiency of light generation due to In segregation and a higher defect density resulting from the lattice mismatch with the substrate [6]. These problems can be overcome by developing novel 2D-3D heterostructures based on GaN interfaced with transition metal dichalcogenides (TMDs) which have lattice constant similar to GaN [7]. The TMDs monolayers have high quantum efficiency of light emission in the red wavelength regime because their direct band gaps are significantly narrower than in GaN. The heterostructure of monolayer molybdenum disulphide (MoS$_2$) interfaced with GaN constitutes an ideal material system to fabricate efficient light emitters over a broad wavelength range covering the lower ultraviolet to the near-infrared range. The interface in such a hybrid-layered material critically influences the electronic transport and carrier transfer. The band-offset, interface phonons, the proximity of defect states of the two material system and the interaction of excitons in the 2D semiconductor with that in bulk III-V semiconductor is expected to modify the carrier mobility and absorption characteristics in addition to the emission properties of the materials.

MoS$_2$ has a strong exciton binding energy in the red region and has multiple excitonic states across the visible to the ultraviolet wavelength range [8]. There is an overlap of the higher energy excitonic absorption state of MoS$_2$ with the absorption domain in GaN. Furthermore, the GaN emission band lies near the Γ-point of MoS$_2$ that have high density of states. The overlap of these electronic states can facilitate energy transfer at the interface between the two semiconductors. The optical properties of monolayer MoS$_2$ are significantly affected by the semiconductor substrate and the relaxation of the excited carriers strongly depends on the levels determines the transport or the relaxation of the optically generated carriers across the interface in the semiconductor heterostructure [12], [13]. The excitation at C exciton state, excitation energy [9]–[11]. The relative position of the energy which is close to Γ-point results in a charge transfer without any momentum change [10], [14]. Optically dressed phonons can significantly influence the dynamics of the excitonic states in MoS$_2$ [15] or graphene [16] based semiconductor nanostructures. In these recent works, the plasmonically induced dressed phonon states were utilized for coherent coupling of excitons and photons. Metal nanoparticles were utilized to drive the phonons for the enhanced light matter interaction that resulted in coherent exciton-plasmon coupling. However, these phonon induced effects can be induced by a polar dielectric surface to influence the excitonic properties of 2D materials.

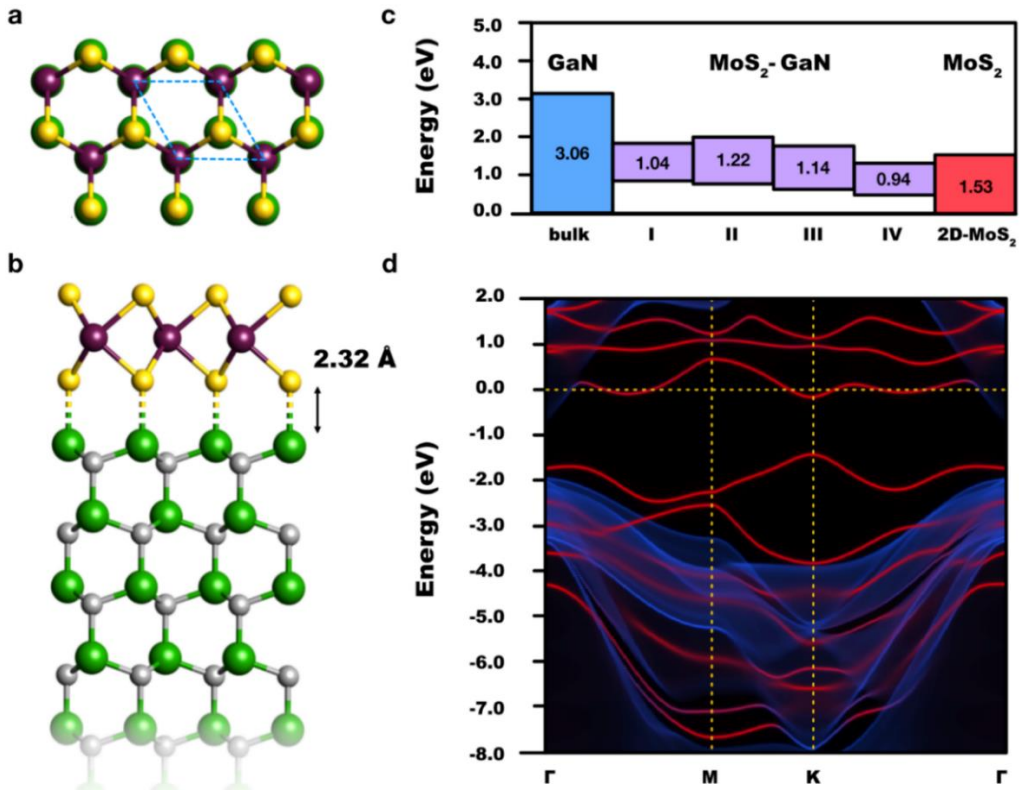

**Fig. 1.** Geometry and electronic structure of MoS$_2$-GaN heterostructure calculated from first principles. Top (a) and side view (b) of the most stable interface structure (II). The unit cell is marked as a blue dashed parallelogram in panel (a). The dashed bonds in (b) denote their mostly van der Waals character. (c) Relative band edge positions of bulk GaN and four interface structures (see supplementary information (SI) for details of geometries I-IV). The bands of isolated MoS$_2$ monolayer are not aligned and the panel indicates only the value of the band gap. (d) Electronic structure of the interface calculated within semi-infinite surface model and projected on MoS$_2$ (red) and GaN states (blue).

GaN is such a polar semiconductor widely used in optoelectronics. Its surface properties strongly depend on the termination (gallium or nitrogen) during the growth. The surface and interface phonon related processes are dominant mechanisms of optically generated carrier relaxation in the nitrides. The relaxation of optically excited carriers is relatively slow in GaN due to hot phonon effects [17]. The electron-phonon interaction in MoS$_2$-GaN heterostructure can modulate the dynamics of the excitonic states in MoS$_2$ [18]. The coupling of phonon modes with the carriers or excitons affects the charge carrier mobility and causes the broadening of the emission bands that is crucial for the development of the broadband lasers and LEDs [19].

In this study, we report the modification of the non-equilibrium absorption characteristics of MoS$_2$ and the photoluminescence (PL) emission properties in 2D MoS$_2$-GaN(0001) vertical heterostructure induced by the interaction of phonon modes with the optically generated carriers. The most stable interface of MoS$_2$-GaN (0001) heterostructure is calculated using a density functional theory (DFT), which revealed a type-I band alignment consistent with our experiments and previously reported results [13], [18]. The valence band maxima and the conduction band minima are almost entirely contributed by MoS$_2$ states. We observed a strong coupling of the optically generated carriers with the phonon modes, transfer of energy across the interface and the capture of carriers in the defect states in the semiconductors. As a result, a significant change in the position, amplitude and the linewidth of the excitonic absorption state of MoS$_2$ at Γ-point as well as an additional fast decay component appears shifted in the presence of the MoS$_2$ layer which is confirmed by the reduction of the absorption calculated by DFT.

## 2. THEORETICAL CALCULATIONS

We performed DFT calculations for van der Waals heterostructures based on MoS$_2$ monolayer and GaN(0001) surface employing the Quantum Espresso package [20], [21]. We used generalized gradient approximation (GGA-PBEsol) in the decay kinetics of excitonic absorption state at K point. The PL emission from GaN is reduced in intensity and red for exchange-correlation functional along with ACBN0 [22], a novel pseudo-hybrid Hubbard density functional approach which ensured the accurate value of the GaN band gap. The ion-electron interaction was treated with the projector augmented-wave pseudopotentials [23] from the pslibrary database [24] while the wave functions were expanded in a plane-wave basis of 80 Ry. The heterostructure was constructed by stacking the $MoS_2$ on the Ga-terminated GaN slab containing 12 Ga-N bilayers and fixing the in-plane lattice constant to the calculated bulk value of the substrate (3.18 Å). The dangling bonds of nitrogen at the bottom side of the slab were passivated by pseudohydrogens. We have considered several stacking configurations and found the most stable one shown in Figure 1 (a-b) with the $MoS_2$-GaN distance of approximately 2.32 Å in line with the previous results [25]. A semi-empirical vdW correction (DFT-D2) [26] was added in order to determine a correct interlayer distance between $MoS_2$ and GaN in each structure, but its value was not taken into account for the analysis of stability. The Brillouin zone sampling at the DFT level was performed following the Monkhorst-Pack scheme using a 10×10×1 k-points grid further increased to 16×16×1 in the projected density of states calculations. The electronic band-structure plots in the form of $MoS_2$ and GaN projected density of states (k, E) maps were obtained using the GREEN package [27]–[29] as a post-processing tool; for these reasons the structure was recalculated self-consistently with the SIESTA code [30] using similar values of all the parameters (e.g. exchange-correlation functional, k-points meshes). The optical properties were calculated employing the PAOFLOW code [31].

## 3. EXPERIMENT

Bulk GaN-monolayer $MoS_2$ vertical heterostructure is synthesized by fabricating monolayer $MoS_2$ layer over the commercially available 4.5 µm thick silicon doped GaN film on a double side polished sapphire substrate (MSE supplies) by CVD method [32]. The equilibrium absorption spectra of the individual semiconductors are measured using a spectrophotometer with a white light source and a photomultiplier detector. Raman characteristics of the heterostructure are studied with a high performance micro-Raman spectrometer equipped with Olympus BX51 microscope using an optical excitation of 2.33 eV.

The non-equilibrium absorption spectrum of $MoS_2$ and the decay kinetics were studied with optical pump-probe spectroscopy. A 100 fs Ti: Sapphire oscillator seeded optical parametric amplifier laser was used as source for pump and probe pulses. The sample is first excited with the pump pulse and the behavior of the excited sample is studied with white-light probe pulses. The difference in absorption of the probe pulses (ΔA) in the presence and absence of pump pulses is measured at different delay times by varying the distance traveled by the pump and probe pulses.

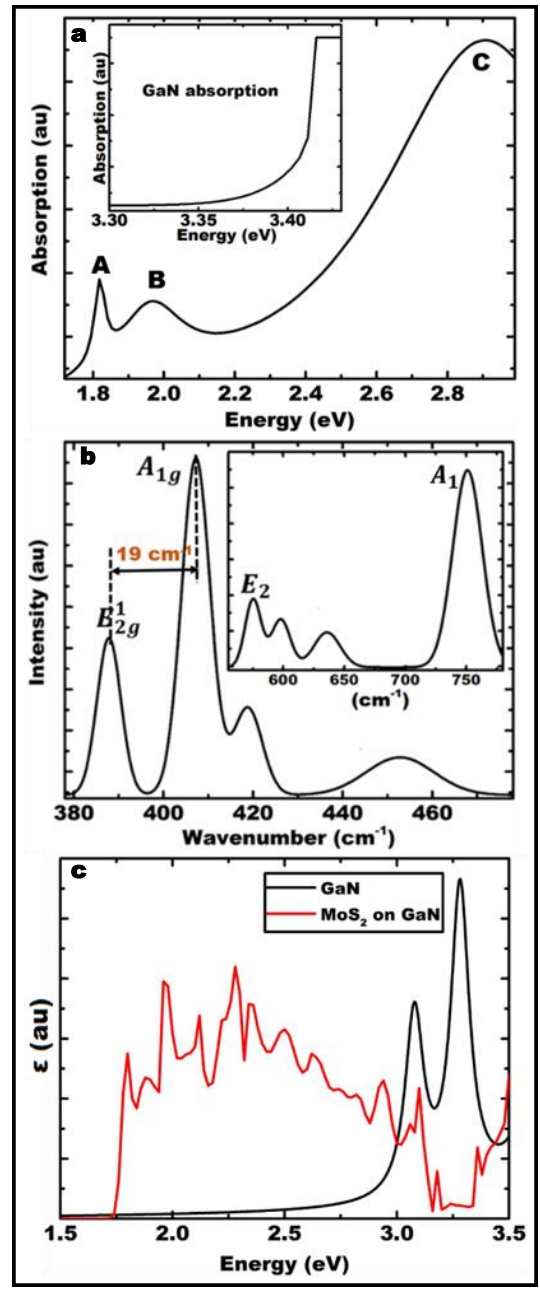

**Fig. 2.** (a) Steady state absorption spectrum of $MoS_2$ on GaN. The excitonic absorption states are located at 1.82 eV, 1.97 eV and 2.91 eV and are denoted by A, B, and C excitons, respectively. The inset shows the absorption band in GaN. (b) Raman characteristics of $MoS_2$ on GaN. The separation between two Raman active modes $E^1_{2g}$ and $A_{1g}$ is 19 cm$^{-1}$ in monolayer $MoS_2$. New Raman modes are activated at different energies. (c) Calculated absorption spectrum showing absorption band in $MoS_2$ and $MoS_2$-GaN.

The PL emission properties of the III-V semiconductor are studied using a home built PL set up. An ultraviolet (UV) excitation line at 3.82 eV from a He-Cd laser source is used to excite the sample. The emitted signal from the sample is collected using a pair of UV collimating lenses in a reflecting geometry. The emitted signal is filtered using a 3.76 eV edge long pass filter and then detected using a CCD spectrometer. The emission characteristics of $MoS_2$ are studied with a home built micro-PL system. The set up consists of a home built upright epifluorescent microscope fitted with 2.25 eV edge dichroic beam splitter. Laser line with an energy of 2.33 eV is focused onto the sample through a 100 X microscope objective. The emitted signal from the sample is collected by the same objective lens, separated from the reflected and the scattered excitation signal passing through a dichroic mirror. The emitted signal from the monolayer $MoS_2$ is further filtered with 2.10 eV edge long pass filter and detected with AD111 photomultiplier tube (PMT) based spectrometer.

## 4. RESULT AND DISCUSSIONS

The optical properties of the heterostructure strongly depend on the structural and electronic details of the interface. The theoretical model based on first principles calculations will be discussed before presenting the experimental observation. The $MoS_2$-GaN (0001) interface assuming the lattice matching between the monolayer $MoS_2$ and the GaN substrate. Several stacking configurations were considered, as shown in the SI, but our calculations clearly favored the structure shown in Figure 1(a-b) with S and Mo atoms aligned with the topmost Ga and N, respectively. The analysis of relative band edge positions (c) indicated that in all the configurations the band gap is strongly reduced with respect to GaN, in agreement with the previous studies[13], [25]. The most stable structure reveals a band gap of predominantly of $MoS_2$ origin, which is further confirmed by the electronic structure (d). Importantly, the type-I band alignment is robust against the structural details; in particular, the band offsets hardly change between four different interface configurations.

The atomic force microscopy (AFM) characteristics show the $MoS_2$ layers with lateral dimensions extending over 5-μm grown on GaN, as shown in SI (Figure S1). The absorption and the Raman spectra provide the static optical characteristics of the heterostructure formed. The steady state absorption spectrum of $MoS_2$ on quartz substrate consists of A, B and C excitonic absorption states at 1.82 eV, 1.97 eV and 2.91 eV respectively, as shown in figure 2 (a). The C excitonic state is located near Γ- point and is identified with large density of states due to band nesting. The inset shows the absorption band in GaN. The UV absorption band in $MoS_2$ overlaps with GaN absorption band [33]. Therefore, by choosing an excitation above 3.4 eV, the carriers in both semiconductors can be simultaneously excited due to interband transition near Γ- point. The active Raman modes in $MoS_2$ - $E_{2g}^1$ and $A_{1g}$ are observed at 388 cm$^{-1}$ and 407 cm$^{-1}$, respectively, as shown in Figure 2 (b). The spacing between these Raman modes combined with the AFM characteristics demonstrate that $MoS_2$ consists of single monolayer. The inset shows the phonon modes in the GaN domain in the heterostructure. The active Raman modes of GaN substrate are shown in SI (Figure S2). The bands centered at 575 cm$^{-1}$ and 738 cm$^{-1}$ represent, respectively, the $E_2$ and $A_1$ longitudinal optical (LO) phonon modes in GaN in the heterostructure. The strong blue shifting of $A_1$ Raman mode of GaN can be explained with the strain induced in GaN layer and the scattering between the active phonon modes in GaN [34]. The enhanced intensity of $A_1$ Raman mode of GaN is attributed to the enhanced coupling of electronic transitions with GaN phonons resulting from the epitaxial growth of $MoS_2$ on GaN. The broad Raman mode centered at 454 cm$^{-1}$ is the combination of second order longitudinal acoustic (2LA) mode and optical mode $A_{2u}$ that represents the coupling between electronic transition and phonons in monolayer $MoS_2$ [35]. The appearance of this Raman mode in the heterostructure indicates the formation of vertical heterojunction [36]. An additional Raman mode is observed in the monolayer at 419 cm$^{-1}$, which has been previously reported as high-order harmonic frequency of an acoustic phonon of GaN and is not an active Raman mode [18], [37], [38]. This Raman mode is coupled with transverse acoustic phonon mode (XA) of $MoS_2$ and generate a new mode at 598 cm$^{-1}$ [39]. Moreover, the additional Raman mode at 636 cm$^{-1}$ is the combined mode of $A_{1g}$ and LA(M) phonon modes in $MoS_2$. The strongly modified Raman characteristics demonstrates a strong interlayer electron-phonon coupling. The absorption spectrum obtained from the DFT simulations shown in Figure 2c predicts that the imaginary part of the dielectric function of the hybrid $MoS_2$-GaN decreases at the GaN band edge.

The effect of the coupling between interfacial phonons and carriers in the transient absorption characteristics of $MoS_2$ are studied with ultrafast optical absorption spectroscopy (Figure 3). The transient absorption spectrum of $MoS_2$ on quartz substrate with an optical excitation of 2.33 eV consists of A, B and C excitonic bands centered at 1.85 eV, 1.99 eV and 2.91 eV, respectively, saturated at a delay time of about 750 fs. The formation of C exciton band with 2.33 eV excitation is attributed to upconversion [40]. The transient absorption spectrum of $MoS_2$ is strongly modified due to GaN substrate. The excitonic bands in $MoS_2$ are formed earlier and saturated faster at a delay time of about 530 fs due to influence of GaN substrate. There is a slight change in the position of A and B excitonic bands; however, the C excitonic band is strongly red shifted to 2.76 eV, as shown in Figure 3(a). The optically excited carriers in $MoS_2$ with the pump pulse are strongly coupled to GaN phonons, which results the changes in the non-equilibrium absorption spectra [18]. The analysis of the decay kinetics shows the enhanced dissociation of the A excitonic band in the presence of GaN substrate, which is consistent with the reported results [18]. A single decay component of 1.07 ps is observed for $MoS_2$ in quartz substrate, which represents the lifetime of A exciton. However, the effective relaxation time of the carriers in $MoS_2$ is reduced on the GaN substrate, and a biexponential decay of excitonic band is observed as represented by the equation; $y = A_1^* exp\left(-\frac{x}{t_1}\right) + A_2^* exp\left(-\frac{x}{t_2}\right) + y_0$. The decay components were observed to be $t_1 = 0.27$ ps and $t_2 = 1.57$ ps, respectively, with the amplitude ratio $A_1^*/A_2^*$ approximately 2:1. The additional fast decay component can be attributed to the electron-phonon scattering induced by the interface phonons in the heterostructure, which leads to exciton dephasing in MoS$_2$.

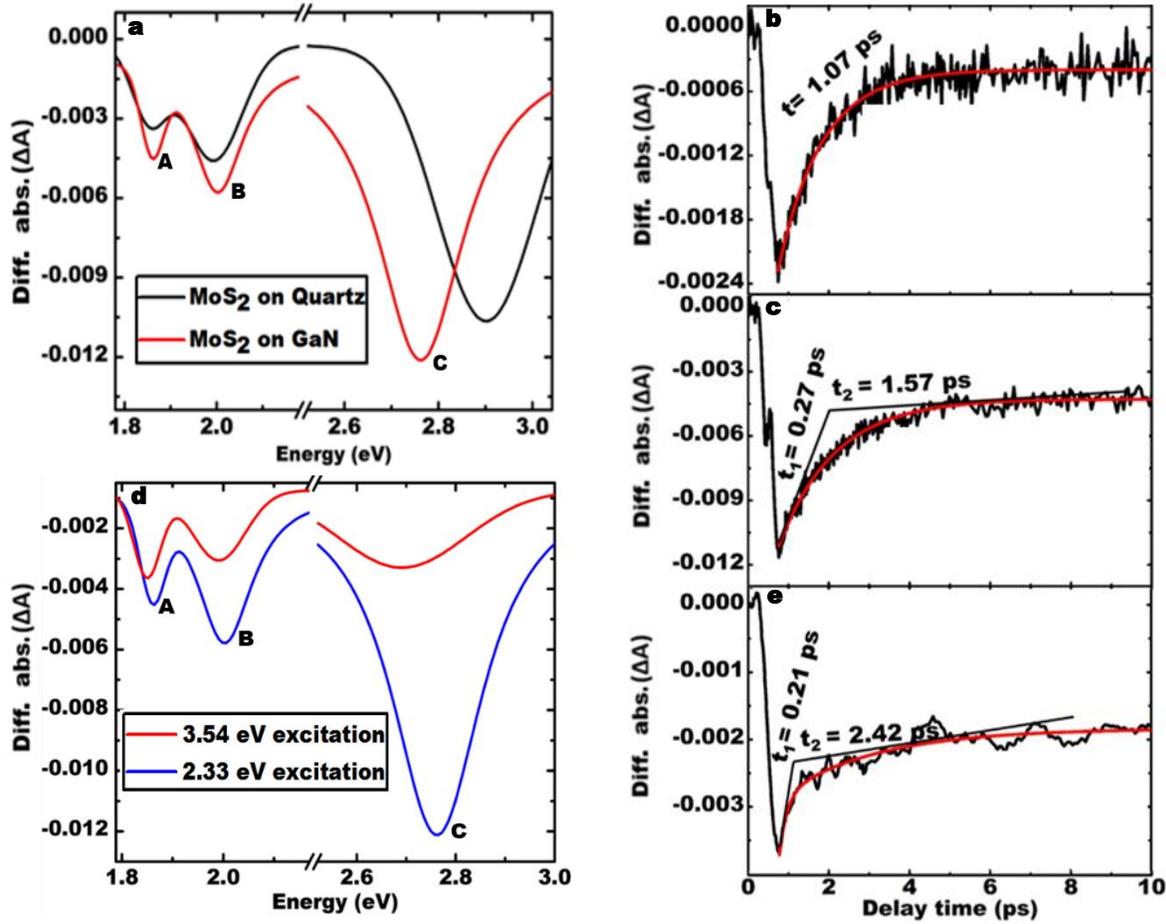

**Fig. 3.** Ultrafast absorption characteristics of monolayer MoS$_2$. (a) Effect of substrate on the transient absorption spectrum measured with 2.33 eV pump excitation. The position of C excitonic absorption state is strongly red shifted in GaN substrate compared to quartz substrate. Decay kinetics of A excitonic band in MoS$_2$ on (b) quartz substrate and (c) GaN substrate. There is single exponential decay on quartz substrate; however, an additional fast decay component appears in the presence of GaN substrate due to carrier capture in GaN layer. (d) Effect of excitation energy on the transient absorption spectrum of MoS$_2$ on GaN substrate. A relative change in intensity and position of A, B and C excitonic absorption states is observed with 3.54 eV excitation. (e) Decay kinetics of A exciton band of MoS$_2$ on GaN substrate with 3.54 eV pump excitation showing the biexponential decay dominated by fast decay component due to carrier capture in both layers.

The effect of excitation energy on the transient absorption spectrum of MoS$_2$ on GaN substrate has also been studied, as depicted in Figure 3(d). Using an excitation pump pulse at 3.54 eV, interband transition of carriers in both MoS$_2$ and GaN is achieved at Γ-point. The drift of optically generated hot electrons in GaN produces non-equilibrium phonon population in momentum space, so-called hot phonons, which slows down the relaxation rate of electrons to the band edges [17], [41]. Thus, the carriers in GaN across the interface are trapped in MoS$_2$ layer, which changes the transient absorption characteristics of MoS$_2$ with 3.54 eV excitation. The A and B excitonic absorption bands are slightly red shifted, but the C excitonic linewidth is broadened and the peak is significantly red shifted to 2.68 eV. The scattering of carriers in MoS$_2$ due to interface phonons causes the spectral broadening of C excitonic band. The decay kinetics of the A excitonic band is dominated by the fast decay component of $t_1 = 0.21$ ps over the slow decay component $t_2 = 2.42$ ps with relative amplitude ratio of 240:1. The type-I band alignment at the MoS$_2$-GaN interface predicted from the DFT calculations facilitates the transfer of the photoexcited electrons in GaN to the conduction band in MoS$_2$ with 3.54 eV excitation. The decrease in transient absorption of the probe pulses owing to the carrier capture process, as shown shown in Figure 3d with 3.54 eV excitation does not occur in case of 2.33 eV excitation. Thus, the relative amplitude of A, B and C excitonic absorption bands changes with 3.54 eV excitation. The carriers transferred to the MoS$_2$ layer relax from its Γ-point to the A and B excitonic states in the presence of the interface phonons and presumably, increases the exciton density in the MoS$_2$-GaN heterostructure compared to the 2.33 eV excitation. The increased scattering of free carriers due to the carrier capture by the defect states in GaN can also contribute to the additional pathways for fast-component of the carrier relaxation [18]. These various relaxation pathways increase the contribution due to the fast decay component as its amplitude is over two orders of magnitude more than that with excitation at 2.33 eV where the photons are not absorbed by GaN. The relaxation rate is also expected to be faster, but was limited by the temporal resolution of our excitation pulse width of the differential transmission spectroscopy measurements.

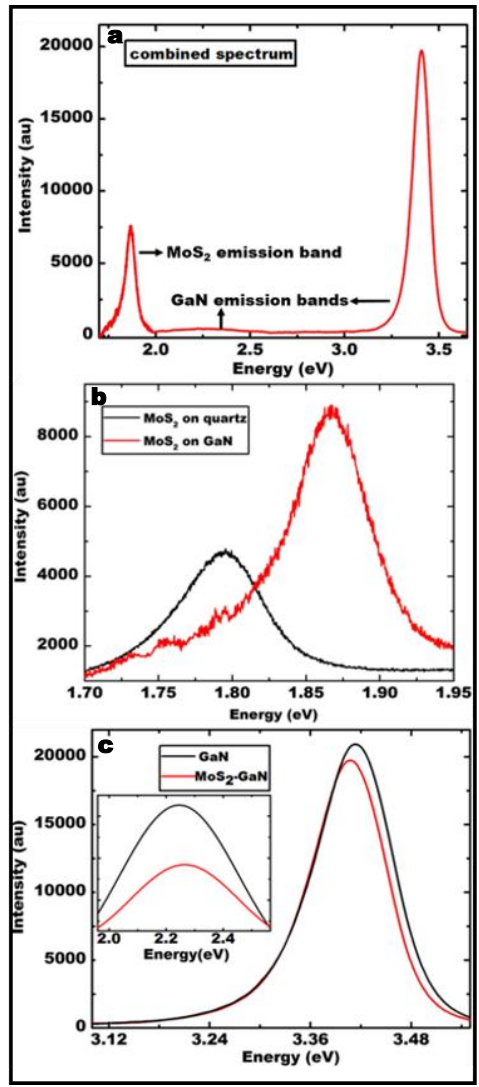

**Fig. 4.** (a) PL emission spectrum from the heterostructure showing the GaN emission centered at 3.41 eV and MoS$_2$ emission centered at 1.87 eV measured with 2.33 eV excitation. (b) Modification of GaN band-edge emission due to heterostructure formation. The inset shows the modified defect band emission after heterostructure formation. (c) Enhanced and blue shifted PL emission spectrum from monolayer MoS$_2$ in GaN substrate compared to quartz substrate.

The PL emission characteristics of the heterostructure are presented in Figure 4. Panel (a) shows the combined emission band of the heterostructure. The emission spectrum consists of GaN emission band centered at 3.41 eV and a weak defect band centered at 2.25 eV measured with 3.82 eV optical excitation and MoS$_2$ emission band centered at 1.87 eV measured with 2.33 eV excitation. The emission spectrum of MoS$_2$ at room temperature consists of excitonic band that corresponds to A exciton recombination and a charged exciton (trion) recombination band, as shown in SI (Figure S3). These emission characteristics are significantly different from the emission spectrum measured in individual semiconductors. The emission spectrum from MoS$_2$ in GaN substrate is significantly changed compared to the emission in quartz substrate, as shown in Figure 4 (b). The emission intensity is enhanced and the emission energy is significantly blue shifted. The contribution of trion in the PL emission is substantially reduced from 47 % to 19 %, the exciton band is blue shifted by 65 meV and slightly broadened, whereas the trion emission band is blue shifted by 21 meV and broadened. In order to explain the modulation of emission properties of MoS$_2$ on GaN substrate, we consider the charge transfer across the interface, as shown in the scheme in Figure 5. As discussed in the previous paragraphs, DFT calculations confirm the type-I band alignment consistent with reported results [13], [18]. With an optical excitation at 2.33 eV, the carriers in MoS$_2$ are excited at K-point at higher energy levels [42]. The carriers tend to relax to A and B excitonic states at K-point in MoS$_2$. However, due to the carrier capture by the defect states in GaN [43], the conversion of trions into excitons in MoS$_2$ results in a significant decrease in the contribution of trions in the PL recombination process. The broadening of the MoS$_2$ emission bands can be related to the coupling of excitonic band with optical phonon modes in GaN [19]. The enhanced and blue shifted PL emission is attributed to the enhanced absorption from the heterostructure and conversion of trions into excitons and the increased lifetime of excitons [9], [13], [43], [44]. The higher blue shift of the exciton band over trion band is attributed to the stronger dependence of emission characteristics of exciton compared to trion [45].

The optical excitation with energy greater than 3.41 eV causes an interband transition of carriers in both semiconductors near Γ-point. The conduction band of MoS$_2$ captures the free-carriers in GaN. These excess carriers in MoS$_2$ can also be captured by the defect states in GaN [9]. Optical excitation in this region enhances the charge transfer between the semiconductors without momentum change [10], [14]. In MoS$_2$ monolayer, carriers are excited to high-energy states due to optical excitation. The optically generated carriers at deep levels, especially holes, have fairly low probability to be scattered to A and B excitonic states at K-point [46]. These carriers are coupled to GaN phonons and lose their energy to crystal lattice, as indicated by the activation of new Raman modes in the heterostructure [18]. Moreover, higher excitation energy increases the probability of intervalley scattering of carriers [44]. The latter minimizes the probability of their relaxation to excitonic state at K-point and hence the radiative recombination is highly reduced. Therefore, no emission is observed in MoS$_2$ with 3.82 eV excitation. On the other hand, the GaN emission band is slightly red shifted, the full width at half-maximum (FWHM) decreases and the intensity of emission is slightly reduced in the heterostructure, as shown in Figure 4(c). The defect band PL intensity from GaN is decreased, but slightly broadened, as shown in the inset. In GaN layer, the optically excited carriers can relax to GaN band edges at Γ-point and recombine to generate PL emission. The formation of heterostructure decreases the absorption cross-section of GaN. The overlapping of the absorption states of MoS$_2$ with the band-edge emission from GaN results in the reabsorption of the GaN emission by MoS$_2$ that decreases the intensity and bandwidth of GaN emission. The slight red shift of emission energy from GaN can be explained by the optically induced band gap renormalization. The increase in bandwidth of the defect band PL emission is attributed to the carrier capture by the defect states in GaN that are transferred from Γ- point of MoS$_2$.

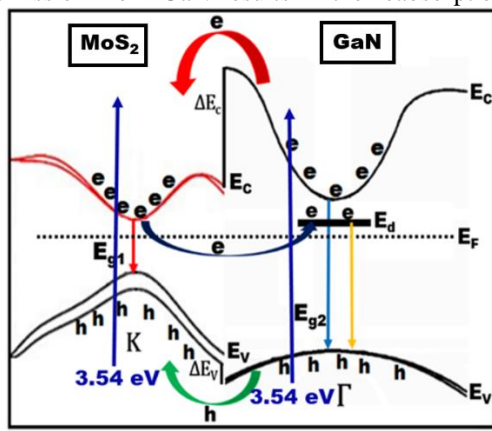

**Fig. 5.** Schematics of the energy band diagram of MoS$_2$-GaN heterostructure showing the band bending, corresponding band offsets and the possible direction of electron and hole transfer across the interface and the radiative recombination in optically excited sample.

## 5. CONCLUSIONS

In summary, we presented the change in the transient absorption characteristics of monolayer MoS$_2$ and the modified PL emission characteristics in monolayer MoS$_2$-GaN (0001) due to the coupling of carriers with the phonon modes and the trapping of carriers at the defect states. The origin and activation of new Raman modes in the heterostructure indicate the electron-phonon coupling between GaN and MoS$_2$. The optical excitation with 2.33 eV causes the interband transition of carriers in MoS$_2$. The capturing of optically generated carriers in MoS$_2$ by the defect states in GaN significantly changes the transient C excitonic band and results in the conversion of trions in MoS$_2$ into excitons that enhances and blue shifts the PL emission from MoS$_2$. Optical excitation with an energy greater than 3.41 eV generates interband transitions in both MoS$_2$ and GaN near Γ- point. Excitation at this energy induces the coupling of carriers in MoS$_2$ with the phonon modes in GaN, interlayer carrier capture and the intervalley scattering in MoS$_2$ that gives rise to change in transient absorption characteristics in MoS$_2$ and modified PL emission in GaN. We believe that our study will be helpful to understand the energy and carrier transfer across the interface is crucial to enhance the device performance in optoelectronic and light-harvesting applications based on MoS$_2$-GaN heterostructure.

**Acknowledgement**:
A.N. and F. D. acknowledge the funding from UNT-AMMPI
JS, PG and MBN acknowledge support by ONR-MURI N000141310635

See Supplement 1 for supporting content.